\documentstyle[aps,graphicx,eqsecnum]{revtex}
\begin{document}
\twocolumn
\def\be{\begin{equation}}
\def\ee{\end{equation}}
\def\bea{\begin{eqnarray}}
\def\eea{\end{eqnarray}}
\newif\ifpdf
\ifx\pdfoutput\undefined
\pdffalse 
\else
\pdfoutput=1 
\pdftrue
\fi

\ifpdf
\usepackage[pdftex]{graphicx}
\else
\fi

\newtheorem{theorem}{Theorem}
\newtheorem{corollary}[theorem]{Corollary}
\newtheorem{definition}{Definition}

\newcommand{\arctanh}{\mathop{\mathrm{arctanh}}\nolimits}
\newcommand{\sech}{\mathop{\mathrm{sech}}\nolimits}

\newcommand{\eqneq}{\!\!\! &=& \!\!\!}


\title{ DYNAMICS OF BLACK HOLE FORMATION IN  AN EXACTLY SOLVABLE MODEL}
\author{Antal JEVICKI
\and 
Jesse THALER}

\address{ Department of Physics, Brown University\\
Providence, Rhode Island  02912  USA}

\renewcommand{\thepage}{}
\maketitle

\begin{abstract}
We consider the  process of black hole formation in particle
collisions in the exactly solvable framework of 2+1 dimensional Anti de 
Sitter gravity.  An
effective Hamiltonian describing the near horizon dynamics of a head on 
collision
is given.  The Hamiltonian exhibits a universal
structure, with a formation of a horizon at a critical distance.  Based 
on it we
evaluate the action for the process
and discuss the semiclassical amplitude for black hole formation.  The 
derived amplitude is
seen to contain no exponential
suppression or enhancement.  Comments on the CFT description of
the process are made.
\end{abstract}

\section{Introduction}

The process of black hole formation in high energy collisions has been
of major theoretical interest [1-4].  Recent scenarios in which the 
Planck mass might
be of the order of a TeV  [5-7]
   have lead to enhanced study of the subject [3-10]. Estimates
for producing black holes at
   LHC were given very recently in  [7-10]. Considerations of black hole 
production in
cosmic ray processes [11-13] were also given.

Simple estimates of black hole formation are based on the hoop conjecture
which results in a geometric
cross section related to the horizon area [15-17].  More exact numerical 
studies
show that black holes indeed
form classically at both zero and nonzero impact parameter collisions 
[18-20].
Nevertheless there have also been
discussions on the possible exponential suppression of the black hole
formation process [21-24].  One is generally interested in describing 
the collision
process in terms of CFT and the corresponding Hilbert space language.
For all the reasons mentioned above, a further analytical study of black 
hole
formation is of certain interest.

The basic quantity of interest is the amplitude for black hole formation
in the collision of two
energetic particles.  At the semiclassical level, the amplitude is given
through the exponential
of the Minkowski action
\be
{\cal A} \sim e^{{i\over \hbar} S{cl}}.
\ee

Even though various studies of classical collisions (and formation of
the black hole horizon)
have been given, the actual amplitude has not been fully evaluated. In
the present paper, we consider the process
in the exactly solvable framework of  2+1 dimensional gravity [30-35].
Using the known  classical
solution, we present a Hamiltonian description of the collision at the
moment when the horizon forms.
This allows for an evaluation of the action  which provides
information on  the question of a possible
exponential suppression (or enhancement) in the formation
process.

The content of the paper is as follows.  In Section 2, we review the 
case of a single particle in black hole background with an emphasis on 
the near horizon dynamics.  In Section 3, we use the known solution of 
the 2-body problem in AdS space-time and describe the time evolution of 
a relative coordinate.  In Section 4, we evaluate the corresponding 
Hamiltonian and the action.  In Section 5, we make comments on the 
differences with the case of a particle in black hole background and 
comparisons with other works.

\section{Particle in a Black Hole Background}

   One could
think of an approximate procedure of estimating the action by 
concentrating on
the relative coordinate describing the distance between the colliding
particles.  In head on collisions, at a critical distance $\, r_H\,$ 
determined by the
total CM energy, a horizon forms and one could then imagine  an analogy 
with  a
case of a particle in a black hole background.
In the case of a particle in a black hole background of fixed mass, the 
well known
semiclassical analysis of Hartle and Hawking [36] provides an
approximation for the amplitude (and the propagator).  An aspect of
the semiclassical calculation  is the
appearance of an imaginary
term in action.

Let us consider the case of
a three dimensional BTZ black hole.
\be
dS^2 \vert_{BH} = - dt^2 \left( {r^2\over l^2} - 8G M \right) +
{dr^2\over {r^2\over l^2} - 8GM} + r^ d \varphi^2
\ee
(This will be the main example considered in our study.)
Concentrating on the radial part
\be
ds^2 = - f (r) dt^2 + {dr^2\over f(r)}
\ee
\be
f(r) = r^2 - r_H^2
\ee
the action for a particle of mass $\, m\,$ is
\be
S = - m \int  \sqrt{f(r) - {\dot{r}^2\over f(r)}} dt.
\ee
It is sufficient in what follows to consider a case of massless
particle whose Hamiltonian is
\be
H = f (r) \vert p \vert.
\ee
The classical value of the action is then given by
\be
S_{cl} = - ET + \int_{r_{0}}^r \, dr \, p (r,E).
\ee
Taking into account the sign ambiguity appropriate for an
ingoing/outgoing particle one has
\be
S_{cl} = - ET \pm E \int_{r_{0}}^r \, {dr\over f (r)}.
\ee
As $r \rightarrow r_H$,
\be
\qquad f (r) \sim 2 r_H \,\, (r-r_H)
\ee
so the classical action diverges. Near the horizon, the  momentum itself
diverges:
\be
p \propto {E\over r - r_H }.
\ee
The velocity on the other hand goes to zero as
\be
\dot{r} = r^2 - r_H^2
\ee
The divergence in the $r$-integration can be avoided by deformation of
the contour [32-34] which results
in imaginary contributions to the action.  More precisely
\bea
\Delta S ( {\rm in}) \eqneq - {i\pi E\over 2r_H}\\
\Delta S ({\rm out}) \eqneq + {i\pi E\over 2 r_H}
\eea

Evaluating the modulus square of the amplitudes, one has
\be
P_{\it emission} = e^{-{E\over T_{H}}} \; P_{\it absorption}
\ee
with the associated Hawking temperature.  The origin of the
the exponential term (and the Hawking effect) is clearly related to the 
divergence of the
classical action (see [36-40] for detailed studies). A more familiar 
appearance of this
effect would be  through a full  Euclidean continuation.

When we consider the process of black hole
formation, we can ask if there is a  similarity of the dynamics involved 
in the
collision with that of a particle in a fixed black hole background. In 
order to  understand that aspect we will construct the relative 
Hamiltonian
for head on collision. It will be seen that even though there are strong
similarities between the two processes (at the level of  equations
of motion), some significant differences will occur at the Hamiltonian 
level and
in the process of  evaluation of the action itself.

\section{Two Particle Collision}
As we have said, we will consider the dynamics of particle collision in 
$2+1$ dimensional
AdS space-time.  AdS$_3$ is given by the constraint
\be
-x_{-1}^2 - x_{0}^2 + x_{1}^2 + x_{2}^2 = -\ell^2
\ee
where $-\ell^2$ is the negative cosmological constant.

Let us first summarize some properties of particle-like and black hole
solutions in this theory.
A point particle of mass $\, m\,$  produces  a
metric

\be
dS^2 = - dt^2 \left( {r^2\over l^2} + \gamma^2 \right) + {dr^2\over
r^2/l^2 + \gamma^2} + r^2 d \varphi^2
\ee
with
\be
\gamma = 1 - 4 G m.
\ee
This represents a cone with the angle
\be
\alpha = 2\pi \gamma
\ee
or a defect angle of
\be
\Delta\alpha = 2\pi (1-\gamma).
\ee

We have that,
\bea
\Delta\alpha = 0 \,\,{\rm for} \,\, m \eqneq 0,\\
\Delta \alpha = 2 \pi\,\,
{\rm for} \,\, m \eqneq {1\over 4G}.
\eea
To obtain the the metric of a BTZ black hole we identify,
\be
\gamma^2 = - 8 GM.
\ee
 From this relation, one sees that a black hole can be obtained for 
imaginary values of $\, \gamma\,$.  Indeed, for a complex value of the 
particle mass such that,
\be \gamma_{BH} = i \sqrt{8GM}
\ee
we have the metric of the BTZ black hole. The process of continuing the
particle mass to complex values is obviously not a physical one but an
identical situation is achieved through collision of two sufficiently
energetic particles.

The dynamics of black hole formation through particle collision in 
anti-de Sitter space has been studied
in detail by Matschull.  In [31] the time dependence of particle 
trajectories was given and the formation
of a black hole horizon was established.  We will use these results to 
construct the relative Hamiltonian
that reproduces the head on  dynamics. For addressing the questions 
raised in the introduction we will
only need the dynamics in the vicinity of the horizon and that is what 
we will concentrate on.
By following the geodesic distance between the two particles, we will
deduce the relative Hamiltonian for the formation.

Let us write
AdS space in terms of cylindrical coordinates $\chi$, $\varphi$, and $t$.
\bea
x_{-1} \eqneq \ell \cosh \chi \cos t \nonumber\\
x_0 \eqneq \ell \cosh \chi \sin t \nonumber\\
x_1 \eqneq \ell \sinh \chi \cos \varphi \nonumber\\
x_2 \eqneq \ell \sinh \chi \sin \varphi
\eea
In these coordinates, the metric is
\be
ds^2 = - \cosh^2 \chi \, dt^2 + d\chi^2 + \sinh^2 \chi \, d\varphi^2
\ee
so $t$ is a time-like coordinate, $\chi$ is radial coordinate, and
$\varphi$ is an angular coordinate.
The collision
of two massless particles studied in [31] is first considered in
the so-called rest frame of the the black hole. One can identify the 
horizon in this frame already,
but the full physical picture of the process is best seen in the 
so-called BTZ frame [34]. We
will make this explicit by performing the transformation of the particle 
trajectories into
BTZ coordinates.

Following [31], the trajectories of two colliding massless particles are 
given    by

\be
\begin{array}{l}
\mbox{{\it Particle 1:}} \hspace{.15 in} \tanh \chi = \sin t
\hspace{.15in} \varphi = 0 \\
\mbox{{\it Particle 2:}} \hspace{.15 in} \tanh \chi = \sin t
\hspace{.15in} \varphi = \theta \hspace{.15in} \sin
\theta = \tanh \mu
\end{array}
\ee

\begin{figure}
\begin{center}
\includegraphics{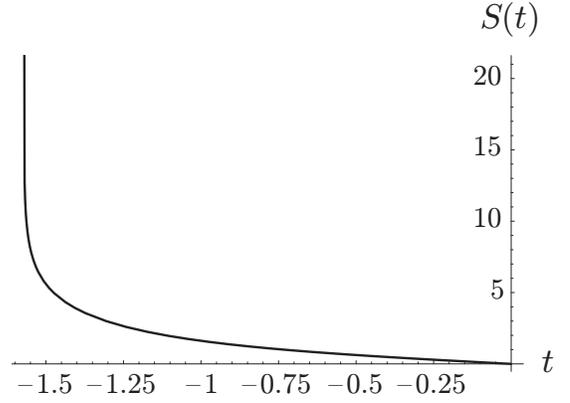}
\end{center}
\caption{Graph of $S(t)$ for $\coth \mu/2 = 2$.}
\label{graph1-fig}
\end{figure}

The energies of the particles are parameterized as
\bea
E_1 \eqneq \tan \epsilon_1 = \coth \frac{\mu}{2} \cosh \mu \\
E_2 \eqneq \tan \epsilon_2 = \coth \frac{\mu}{2}
\eea
and $\mu$ will represent  the half length of the black hole horizon.
The geodesic that connects the two particles is described by
\be
\tanh \chi \sin (\epsilon_2 + \varphi) = - \sin t \sin \epsilon_2.\ee
Because the part of the geodesic for which we are concerned runs from
$\varphi = 0$ to $\varphi = \theta$,
it is convenient to parameterize the geodesic in terms of $\varphi$.
Thus, the geodesic length will be
\be
S(t) = \int_0^\theta d\varphi \;
\sqrt{\left(\frac{d\chi}{d\varphi}\right)^2 + \sinh^2 \chi} .
\ee
To find $\frac{d\chi}{d\varphi}$ we can differentiate the equation
describing the geodesic implicitly
with respect to $\varphi$.  Substituting this into the integral, we find
the
following geodesic distance:
\be
S(t) = -2 \; \arctanh \left(  \frac{\sin t}{\sqrt{1+\cos^2 t \coth
\mu/2}} \right)
\ee
Time runs from $t=-\pi/2$ to $t=0$ and we can see from Figure
\ref{graph1-fig} that as $t$
goes to $0$, $S(t)$ smoothly approaches $0$.

We will now perform the change to the BTZ coordinate. It is in that 
coordinate
system that the formation of the horizon can be explicitly seen. We 
follow [34] and
denote the metric (which is the same as the usual BTZ  metric) as
\be
ds^2 = -\left(\frac{\tilde{r}^2}{\ell^2} - 8GM\right) d\tilde{t}^{\,2} +
\frac{d\tilde{r}^2}{\frac{\tilde{r}^2}{\ell^2} - 8GM} + \tilde{r}^2
d\tilde{\phi}^2.
\ee
with the tilde introduced to avoid confusion with a fixed black hole 
case.
Consider a change from the
coordinates $\tilde{r}$, $\tilde{\phi}$, and $\tilde{t}$,
to new coordinates
$R$, $\Phi$, and $T$ given by
\be R = \frac{\tilde{r}}{\mu}, \hspace{.3 in} \Phi =  \mu \tilde{\phi},
\hspace{.3 in} T =
\mu \left(\tilde{t} + \frac{\pi}{2}\right).
\ee
For the region outside of the horizon $\tilde{r} = \mu$,
the range of these coordinates are
\be 1 < R < \infty, \hspace{.3 in} -\mu < \Phi < \mu, \hspace{.3 in}
\frac{\mu\pi}{2} < T < \infty.
\ee
The metric is now
\be
ds^2 = -(R^2 -1)dT^2 + \frac{dR^2}{R^2 -1} + R^2 d\Phi^2.
\ee

This exhibits the connection between the BTZ black hole and the AdS 
space itself.  In fact, in terms
of the $x_i$ variables:
\bea
x_{-1} \eqneq \sqrt{R^2 -1} \sinh T \nonumber\\
x_0 \eqneq R \cosh \Phi \nonumber\\
x_1 \eqneq \sqrt{R^2 -1} \cosh T \nonumber\\
x_2 \eqneq R \sinh \Phi
\eea
As a consequence, one has a relationship between the cylindrical AdS 
coordinates with the
BTZ coordinates.  The trajectories of the particles in the new frame read
\be
\begin{array}{l}
\mbox{{\it Particle 1:}} \hspace{.3 in} R = \coth T \hspace{.3in} \Phi =
0 \\
\mbox{{\it Particle 2:}} \hspace{.3 in} R = \coth T \hspace{.3in} \Phi =
\mu
\end{array}
\ee

\begin{figure}
\begin{center}
\includegraphics{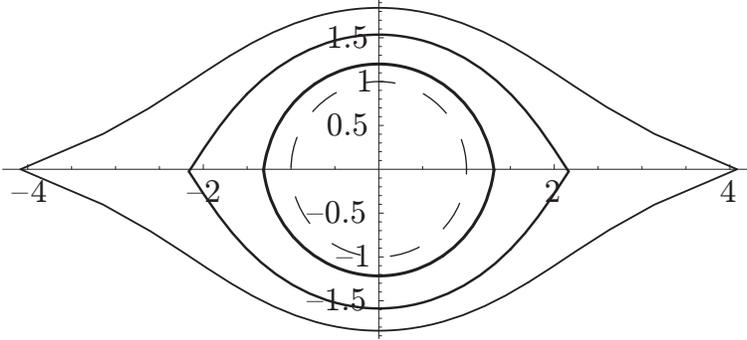}
\end{center}
\caption{Graph of BTZ geodesic for $w = 2$, and $T = .25, .50, 1.00$.}
\label{graph3-fig}
\end{figure}

The geodesic connecting the particles in the BTZ coordinates is given by
\be
R^2 = \frac{w^2 \cosh^2 T}{w^2 \cosh^2 T - (\sinh \Phi + w \cosh
\Phi)^2}
\ee
where we have introduced the parameter $w = \coth \mu/2$.  (This is both
the geodesic obtained by solving the geodesic equation and the
transformation of the geodesic from AdS space.)

In figure \ref{graph3-fig}, we have plotted the geodesic for various
times $T$ using $R$ as a radial coordinate and $\Phi$ as an angular
coordinate.  The dashed line represents the horizon of the black hole at
$R=1$.  We can see from this figure that at $T \to \infty$, the
particles get closer to the black hole horizon but never enter.  Because
$\Phi$ runs from $-\mu$ to $\mu$, the length of the horizon is $2\mu$,
so $\mu$ has the same meaning in our BTZ coordinates as in AdS space.

\begin{figure}
\begin{center}
\includegraphics{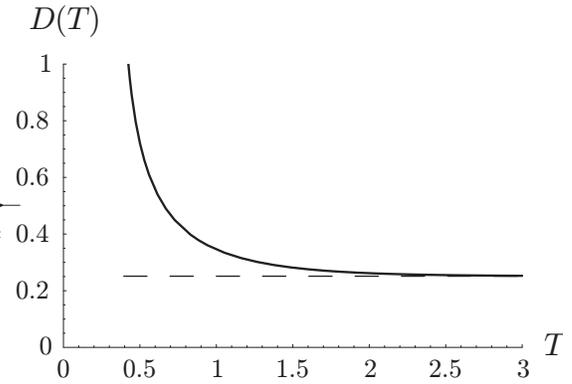}
\end{center}
\caption{Graph of $D(T)$ for $w = 8$.}
\label{graph2-fig}
\end{figure}

We are now ready to derive the geodesic distance in the BTZ metric.  In
analogy with the AdS
case, it is convenient to parameterize the geodesic in terms of the
variable $\Phi$.  The spacial geodesic distance is given by
\be
D(T) = \int_0^\mu d\Phi \; \sqrt{\frac{1}{R^2 -1}\left(\frac{dR}{d\Phi}
\right)^2 + R^2}.
\ee
Again using implicit differentiation to find $\frac{dR}{d\Phi}$, we find
the relative distance to be
\bea
D(T) \eqneq \arctanh \left( \frac{3 + w^2 + 2w^2 \cosh^2 T}{(1+w^2)
\cosh T \sqrt{1+w^2 \sinh^2 T}}
\right) \nonumber \\
&& \mbox{         } - \arctanh \left( \frac{1}{\cosh T \sqrt{1+w^2
\sinh^2 T}}  \right).
\eea
As we can see from Figure \ref{graph2-fig}, $D(T)$ approaches the
limiting value of $\mu$ as $T \to \infty$.

\section{The Hamiltonian and the Action}

In the above sections, we have constructed an exact solution for the
time evolution of our particles in the BTZ coordinates.
  From this, we will now deduce the Hamiltonian describing the relative
dynamics of the particles as they approach the near horizon limit.

The Hamiltonian is to reproduce the equation of motion through
\be
\dot{D} = \frac{\partial H}{\partial P_D}.
\ee
Let us therefore first find the appropriate expression for $\dot{D}$ in
the limit of $D \to \mu$
or equivalently $T \to \infty$.

We can expand $D(T)$ in powers of $e^T$ about the point $T=\infty$.
\bea
D(T) &=& \arctanh\left(\frac{2w}{1+w^2}\right) + \left
( \frac{4(1+6w^2+w^4)}{w(w^2-1)^2}\right) e^{-2T}\nonumber\\
&&+ O(e^{-4T}).
\eea
Recalling that $w = \coth \mu/2$, we have:
\be
D(T) \approx \mu + 4\cosh 2\mu \tanh \frac{\mu}{2}e^{-2T}
\ee
\be
\dot{D}(T) \approx -8\cosh 2\mu \tanh \frac{\mu}{2}e^{-2T}
\ee
As $T \to \infty$, we have the relation (up to order $e^{-2T}$)
\be
\dot{D} = \frac{\mu^2 - D^2}{\mu}.
\ee
This result looks much like the case of particle falling into a black
hole background as given by eq.(11). As
we will now show, the Hamiltonian is very different.

We know that the mass of the (effective) black hole is the total energy
of the system.  For the BTZ
black hole, $\mu = \ell \sqrt{8GM}$ so in units where $8G\ell^2 = 1$,
the Hamiltonian $H = \mu^2$. It is the fact that in the collision case, 
the black hole mass is given by the Hamiltonian itself which leads to a 
difference from the situation in the case of a single particle.  Now the 
horizon radius is a dynamical variable  (related to the Hamiltonian).  
The Hamiltonian is consequently to be self consistently determined.

We can reproduce the above equation taking a Hamiltonian of the form
\be
H = D^2 \tanh^2 \frac{P}{2D}.
\ee
In the limit $D^2 \to \mu^2  = H$, this Hamiltonian will give the
desired relation for $\dot{D}$.  Indeed
\be
\dot{D} = \frac{\partial H}{\partial P}  = \frac{\sqrt{H}}{D^2} (H - 
D^2) \approx
\frac{H - D^2}{\sqrt{H}}
\ee

The relative Hamiltonian clearly exhibits the formation of the horizon.  
Consider the Hamiltonian for large momenta $\, P\,$, when it takes the 
form:
\be
H - D^2 = - 4D^2 \, e^{-{P\over D}}
\ee
We see that for distances $\, D \sim H^{1/2} $, one has a typical 
behavior associated with an horizon of a black hole. We comment that eq.(48) can also be taken as giving the Hamiltonian (one should remember that we are working in the near horizon limit). Let us study  
 the growth of momenta near the horizon.  Even though the equation of 
motion for the velocity
\be \dot{D} \propto D^2 - H
\ee
appears to be the same as in the particle case, the behavior of the 
momentum is very different.  We have
\be
P \approx - D \ln \, {D^2 - H\over 4D^2}
\ee
Compared with the particle case, where the momentum was diverging 
linearly (eq. 10), we now have a milder logarithmic divergence.

Now we can evaluate the action for the classical process of black hole
formation.  Apart from the energy
term responsible for energy conservation, we have
\be
\bar{S} = \int^\mu P dD.
\ee
We evaluate this quantity for our effective Hamiltonian.  We have
\be
P = -2D \arctanh \frac{\mu}{D}
\ee
The following change of variables will be helpful for evaluating the 
integral:
\be
z = \frac{\mu}{D}  \hspace{.3in} dz = -dD\frac{\mu}{D^2}  \hspace{.3in}
dD =-dz \frac{\mu}{z^2}
\ee
We find
\bea
\int^\mu PdD \eqneq 2\mu^2 \int^1 \frac{dz}{z^3} \arctanh z \nonumber \\
\eqneq \left. 2\mu^2 \left(  \frac{1}{2}\left(1-
   \frac{1}{z^2}\right) \arctanh z - \frac{1}{2z}\right) \right|^1\\
&& = 
-\mu^2\nonumber
\eea
and thus no divergence from near the horizon.
  So we have the  result that the action for the classical process is
finite and it does not have imaginary contribution.  As a consequence, 
the amplitude is a pure phase.
\be
{\cal A}  \propto \, e^{{i\over \hbar} S_{ce}} = e^{{i\over \hbar} 
\Delta}
\ee
  This implies that of the semiclassical level, there   is no exponential 
suppression (or
enhancement) of this process.

\section{Comments }
In the main body of this paper we have described the near horizon 
dynamics
of the two body problem in AdS gravity in terms of an effective 
Hamiltonian.
Even though the Hamiltonian is derived in the particular framework of 
$2+1$
dimensional Anti de Sitter gravity we believe that it contains some 
universal
features which will be there in any dimensions.  We have seen that this
effective Hamiltonian predicts a particular logarithmic growth of the 
relative
momentum in the near horizon limit.  As we have emphasized this 
represents a
difference from the case of a particle in a fixed mass black hole 
background.
The effective Hamiltonian that we are lead to differs from some other
approaches, for example [23].

As a consequence of a softer, logarithmic growth of the near horizon
momentum there appears no divergence when integrating to the action.  The
action for the proces is seen to be finite implying a pure phase 
contribution to the
semiclassical amplitude. Consequently we do not find in this analysis
a possibility for an imaginary contribution or exponential suppression
of the kind advocated in [21,24].

Concerning the form of the effective Hamiltonian, we would like to 
comment on
one other relevant issue. Its form exhibits a higher nonlinearity in
the momentum (and relative coordinate). In that sense we have a 
similarity
with studies of Unruh [41] and Corley and Jacobson [42] where 
modifications
to the energy momentum dispersion were considered.
  It was shown that  that  nonlinearities
introduced do  not modify the Hawking effect at least at low energies. 
In our case,
we have seen that the nonlinearities of the effective Hamiltonian
\emph{do} modify the near horizon growth of the momentum.  One
should remember that in the case of a black hole formation, one deals
with high energies and consequently there is no obvious discrepancy
with the findings of [42].

Our final comment concerns the very interesting problem of formulating
the process of black hole formation in a conformal field theory (CFT)
framework. This question is of major interest and the main trust of
the AdS/CFT correspondence (see [33,34] for example and the corresponding
references). The relative Hamiltonian that we have presented can be
of direct relevance for understanding this issue.  In particular it
should be possible to  obtained this Hamiltonian from CFT. A  way in 
which we can see
this happening is by extracting the corresponding ``collective''
degrees of freedom [43] from the analogue CFT.  Results
along this line will be given in [44].
\vskip .25in

\noindent{{\bf Acknowledgements}

We would like to acknowledge helpful
discussions with Robert De Mello Koch,
Joao Rodrigues, Lenny Susskind and David Lowe.
We would in particular like to
thank Sumit Das for his help and Steven Corley for
clarifications concerning the
role of high frequency modes. A. J. would like to thank Prof. J. Rodrigues for his hospitality at the University of Witwatersrand and Prof. Hendrik Geyer for his hospitality at the Stellenbosch Institute for Advanced Study, South Africa, where part of this work was done.
\\

This work is supported in part by Department of
   Energy Grant \# DE-FG02-91ER40688 - Task A.

\end{document}